# A wind, temperature, H$_2$O and CO$_2$ scanning lidar mobile observatory for a 3D thermodynamic view of the atmosphere

Fabien Gibert[1], Dimitri Edouart[1], Paul Monnier[1], Claire Cénac[1], Vincent Gauthier[1], Henri Salvador[1]

[1] Laboratoire de Météorologie Dynamique (LMD/IPSL), École Polytechnique, Institut Polytechnique de Paris, Sorbonne Université, École normale supérieure, PSL Research University, CNRS, École des Ponts, Palaiseau, France, E-mail : gibert@lmd.polytechnique.fr

**Abstract.** A ground-based mobile 3D lidar observatory has been developed for simultaneous measurements of wind speed, temperature, water vapor and carbon dioxide absorption in the atmosphere. The present paper reports details of the instruments, assesses the current performances and gives some examples of measurements for different geophysical applications.

## 1 Introduction

The motivation of this work is to provide advanced observations of the main variables that characterize the land-atmosphere exchanges of momentum, temperature, water vapor and carbon dioxide. The lidar mobile observatory is a prototype that hopefully will help to path the way to a future 3-D thermodynamic view of the atmosphere that will match current and future Navier-Stokes equation simulation at different scales (Large-Eddy Simulation (LES), Direct Numerical Simulation (DNS)) and will help our understanding of the carbon cycle in the context of global warming [1]. Multiple goals include: (i) to address the representativeness of in situ measurements in heterogeneous landscape, especially for surface fluxes (ii) to assess the relevance of Monin-Obukhov similarity theory (MOST) which links gradient and flux close to the surface (iii) to address the issue of dissimilarity of scalar transport such as heat and water vapor or CO$_2$ in inhomogeneous landscape (iv) to help to find advanced model parametrizations of land-surface or boundary layer - free atmosphere exchanges and transport processes, for both convective and stable planetary boundary layers. To do so, new observations are needed that can provide, first, a 3-D view of the atmosphere and second, that have turbulence-scale temporal and spatial resolutions in order to investigate flux-gradient relationships and estimate higher-order moments.

In this paper we present the main characteristics of the two lidars of this observatory and we assess the performances in terms of precision and bias.

## 2    Instrumental set-up

### 2.1    3-D lidar observatory

The mobile 3-D lidar observatory consists in two containers with scanning lidars that operates in non-visible UV and the NIR optical windows for eye safe reasons (Fig. 1). First lidar is a temperature and water vapor Raman lidar at 355 nm (TERA) and second is a prototype DIAL and Doppler lidar at 2051 nm (COWI) for simultaneous wind speed and carbon dioxide ($CO_2$) absorption measurements. Lidar observations are completed by several in situ sensors to complete the dataset especially in the surface atmospheric layer. Two in situ flux stations with sonic anemometers and gas analyzers, one attached to the observatory and second deployed few kilometers apart are especially used for turbulent-linked measurement references and to assess surface property heterogeneities.

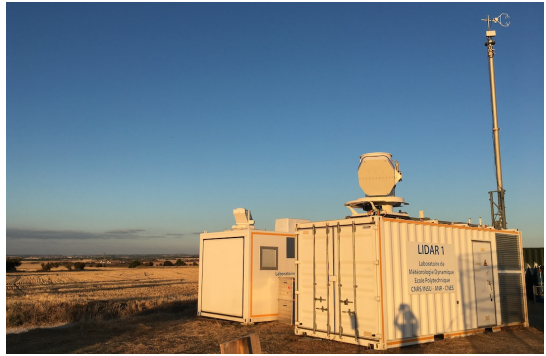

**Fig. 1.** 3-D lidar observatory in operation during the LIAISE field experiment, July 2021, Preixana, Spain.

### 2.2    TERA: temperature and water vapor Raman lidar

The main characteristics of TERA lidar are shown in Fig. 2. The lidar consists in a diode-pumped and seeded tripled Nd:YAG laser that provides 200 mJ pulses at 100 Hz at 354.8 nm (Merion-Lumibird SA) and a 50 cm diameter telescope and associated scanning device (Fig. 2). The detection set-up includes several interference filters in cascade, two for temperature rotational channels (RR1- 354.15 nm (0.3 nm bandwidth) and RR2 - 353.3 nm (0.5 nm)), one for H2O vibrational Raman detection at 407.7 nm (0.3 nm) similar to the system that was developed in [2]. An elastic channel is used to monitor the structure of the atmosphere but also to calibrate the 3-D axes of the scanning device with referenced hard targets. The detection and acquisition system use standard devices with PMTs and LICEL TR40-12-bit systems for simultaneous analogic and photon-counting detections. The power supply, including the cooling of the laser is lower than 5 kW. The scanning device has been built in the lab and uses custom ellipsoidal gas fusion mirrors with an aluminum coating. TERA was specifically designed for high temporal and space resolution profiling of temperature and water vapor in the boundary layer for turbulence-linked measurements but longer averaging scales (1 h - 200 m) enables to make measurement up to the stratosphere during the night.

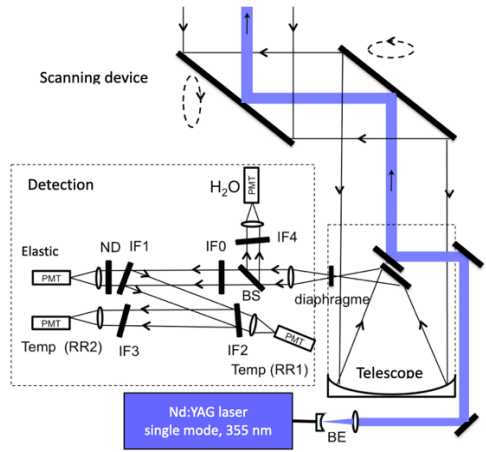

**Fig. 2.** TERA optical set-up.

### 2.3 COWI: $CO_2$ and wind lidar, Doppler and DIAL

COWI lidar makes use of a Thulium fiber laser pumped dual wavelength seeded Ho:YLF MOPA emitter that provides 10 mJ pulses at 2 kHz [3].

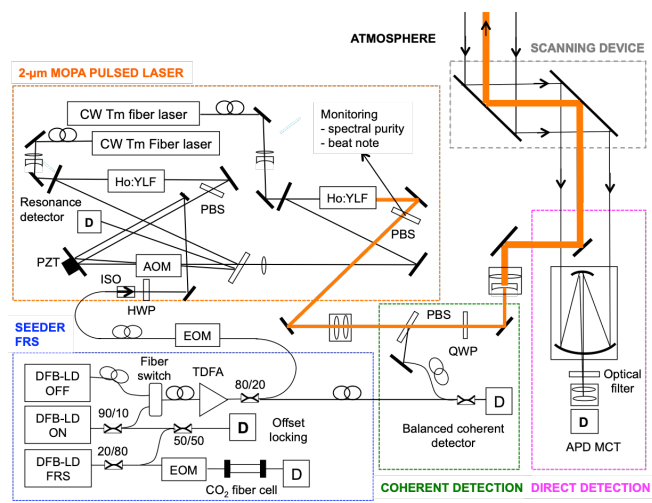

**Fig. 3.** COWI optical set-up.

The lidar has a coherent detection for wind speed measurement and a direct detection using a new HgCdTe APD and a 20 cm diameter telescope for differential absorption measurement of $CO_2$ [4]. In the present configuration, two wavelengths are used to make DIAL measurements of $CO_2$ using the R30 absorption line at 2050.97 nm but a third one may be added at 2050.53 nm to make simultaneous measurement of $H_2O$. Spectral purity is measured to be larger than 99.96% and frequency stability (better than 150 kHz at 10s) is achieved using a lab-made frequency reference system that relies on a CO2-filled absorption cell, external frequency modulation and Pound-Drever-Hall technique. Total power supply of the lidar is lower than 2 kW.

## 3   Measurements and performances

### 3.1   3-D lidar sensing

The main objectives of 3-D lidar sensing are to provide tropospheric vertical profiles and to document at the same time the heterogeneity of surface layer properties that could explain these vertical profiles (Fig. 3).

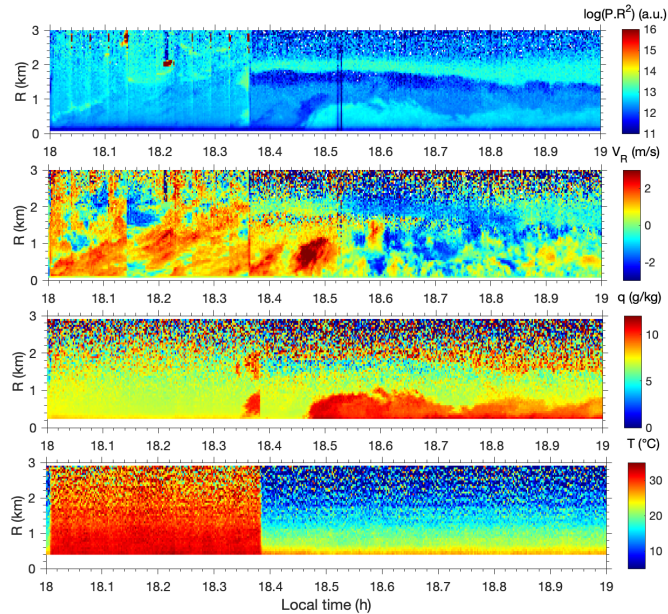

**Fig. 4.** Typical scanning lidar observations during one hour. Top to bottom: lidar reflectivity at 2 μm, radial wind speed, specific humidity and temperature. Measurement starts with four 6° RHI, then a 30° PPI, again four 6° RHI and vertical measurements for the rest of the hour. Time en range resolution are 8 s and 50 m respectively.

In this way, the lidars are operated sequentially in three different modes: (i) vertical to get the troposphere characteristics, scalar profiles and moments, integral scales and fluxes using the eddy-covariance method; (ii) RHI (range-height indicator), low altitude vertical cross-section of the surface layer at a given azimuth angle to apply Monin-Obukhov similarity theory and estimate surface fluxes heterogeneity; (iii) PPI (plan polar indicator), low altitude horizontal cross-section of the surface layer to measure scalar field heterogeneity.

### 3.2    Potential temperature and specific humidity gradients and fluxes

One of the main objectives of this lidar observatory is to provide a continuous thermodynamic view of the convective boundary layer in order to improve model parametrizations of vertical transport. Correlation between radial wind speed and scalar profiles (temperature, $H_2O$, $CO_2$) requires co-located instruments, several altitude-referenced targets to calibrate scanning device elevation and azimuth angles and synchronized data acquisition systems. Fig. 4 shows an example of eddy-covariance sensible and latent heat fluxes calculated with time and space resolution of 30 min and 50 m. Instrumental and sampling errors are indicated. Mean potential temperature and specific humidity profiles are also displayed. Statistical errors are respectively lower than 0.5 K and 0.3 g/kg with time and space resolution of 2 min and 7.5 m in the first kilometer of the CBL. Comparison with radiosondes profiles show a bias for potential temperature profile for $z < 0.3$ km due to a different overlap function for the two lidar temperature channels.

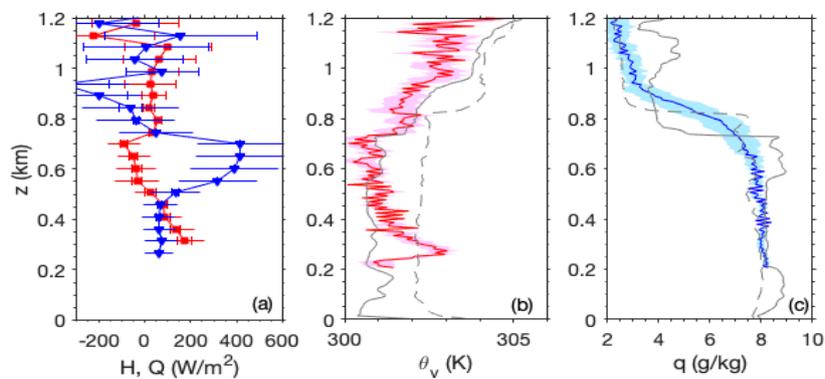

**Fig. 5.** (a) Lidar eddy covariance sensible heat (H) (red) and latent heat (Q) (bleu) flux profiles between 13.4 and 14 h calculated with native data resolution of 50 m and 8 s. Long cap error bars are for the instrumental error and no cap error bars for sampling error. (b) Virtual potential temperature profile and standard deviation (c) Specific humidity profile and standard deviation Time and space resolution are 2 min and 7.5 m. Radiosondes profiles are indicated at 13 h (solid grey line) and at 14 h (dashed grey line).

### 3.3 Some insight in $CO_2$ DIAL measurements

Simultaneous measurements of $CO_2$ with $H_2O$ and temperature is of great interest to address the issue scalar diffusivity difference and scalar dissimilarity in the boundary layer. Up to now, such measurement is still difficult to achieve given the required precision and accuracy (< 1%) especially when one wants to study turbulent exchanges. At least the $CO_2$ field heterogeneity may be investigated with a useful geophysical precision but still limited time and space resolution [5]. Fig. 6 shows an example of $CO_2$ diurnal cycle monitoring both in the surface layer and in the boundary-layer using the scanning ability of the COWI lidar. Measurements were made using the coherent detection. Unfortunately, the direct detection that uses the HgCdTe APD prototype detector has not been successful at this point [4]. A new detector made in the framework of the ongoing HOLDON H2020 [6] project is expected to give the necessary precision, space and time resolution for serious $CO_2$ geophysical studies.

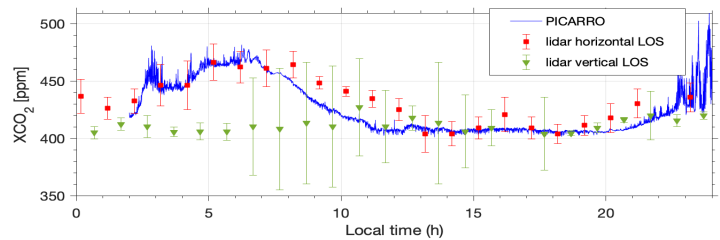

**Fig. 5.** $CO_2$ coherent differential absorption lidar measurements with time and space resolution of 30 min and 500 m. PICARRO: in situ gas analyser.

**Acknowledgements**

This work is supported by the European Space Agency (ESA), the National Space Agency (CNES), the Directorate General of Armaments (DGA), the National Centre for Scientific Research (CNRS) and the Research National Agency (ANR).